\documentclass[slac_one]{revtex4}
\usepackage{graphicx}
\usepackage{fancyhdr}
\def\be{\begin{equation}}
\def\ee{\end{equation}}
\def\bc{\begin{center}}
\def\ec{\end{center}}
\def\bea{\begin{eqnarray}}
\def\eea{\end{eqnarray}}

\def\gappeq{\mathrel{\rlap {\raise.5ex\hbox{$>$}} {\lower.5ex\hbox{$\sim$}}}}
\def\lappeq{\mathrel{\rlap{\raise.5ex\hbox{$<$}} {\lower.5ex\hbox{$\sim$}}}}
\def\MeV{{\rm MeV}}\def\GeV{{\rm GeV}}

\def\swsqeffl{\sin^2{\theta_{eff}}}
\def\dalhad{\Delta\alpha^{(5)}_{had}(m_Z)}

\begin{document}

\hfill{~~~~~~~~~~~RM3-TH/06-21}

\hfill{~~~~~~~~~~~CERN-PH-TH/2006-228}

\vskip 1cm

\title{Introduction to the Terascale} 

%

\author{Guido Altarelli}
\affiliation{Dipartimento di Fisica `E.~Amaldi', Universit\`a di Roma Tre
and INFN, Sezione di Roma Tre, I-00146 Rome, Italy and \\ CERN, Department of Physics, Theory Division, 
 CH--1211 Geneva 23, Switzerland}

\begin{abstract}
We briefly review the status of the electroweak theory, in the Standard Model and beyond, on the brink of the LHC start
\end{abstract}

\maketitle

\thispagestyle{fancy}



\section{THE PROGRAMME OF LHC PHYSICS}
The first collisions at the LHC are expected near the end of '07. The physics run at $14$ TeV will start in the spring of '08. The particle physics community eagerly waits for the answers to the big questions that one expects to get from the LHC. The top physics issues at the LHC, addressed by the ATLAS and CMS collaborations, will be: 1) the experimental clarification of the Higgs sector of the electroweak (EW) theory, 2) the search for new physics at the weak scale that, on conceptual grounds, one predicts should be in the LHC discovery range, and 3) the identification of the particle(s) that make the dark matter in the Universe. In addition the LHCb detector will be devoted to the study of precision B physics, with the aim of going deeper in the knowledge of the Cabibbo-Kobayashi-Maskawa (CKM) matrix and of CP violation. The LHC will also devote a number of runs to accelerate heavy ions and the ALICE collaboration will study their collisions for an experimental exploration of the QCD phase diagram.

\section{THE HIGGS PROBLEM}

The Higgs problem is really central in particle physics today. On the one hand, the experimental verification of the Standard Model (SM) cannot be considered complete until the structure of the  Higgs sector is not established by experiment. On the other hand, the Higgs is also related to most of the major problems of particle physics, like the flavour problem and the hierarchy problem, which in turn strongly suggest the need for new physics near the weak scale. In turn the discovery of new physics could clarify  the dark matter identity. It is clear that the fact that some sort of Higgs mechanism is at work has already been established. The W or the Z with longitudinal polarization that we observe are not present in an unbroken gauge theory (massless spin-1 particles, like the photon, are transversely polarized). The longitudinal degree of freedom for the W or the Z is borrowed from the Higgs sector and is an evidence for it. Also, the couplings of quarks and leptons to
the weak gauge bosons W$^{\pm}$ and Z are indeed precisely those
prescribed by the gauge symmetry. The accuracy of a few per mil in
the precision tests implies that, not only the tree level, but also the
structure of quantum corrections has been verified. To a lesser
accuracy the triple gauge vertices $\gamma$WW and ZWW have also
been found in agreement with the specific predictions of the
$SU(2)\bigotimes U(1)$ gauge theory. This means that it has been
verified that the gauge symmetry is unbroken in the vertices of the
theory: all currents and charges are indeed symmetric. Yet there is obvious
evidence that the symmetry is instead badly broken in the
masses. Not only the W and the Z have large masses, but the large splitting of, for example,  the t-b doublet shows that even a global weak SU(2) is not at all respected by the fermion spectrum. This is a clear signal of spontaneous
symmetry breaking and the implementation of spontaneous
symmetry breaking in a gauge theory is via the Higgs mechanism. The big remaining questions are about
the nature and the properties of the Higgs particle(s). The present experimental information on the Higgs sector, mainly obtained from LEP as described in section 4, 
is surprisingly limited. It can be summarized in a few lines, as follows. First, the relation $M_W^2=M_Z^2\cos^2{\theta_W}$, modified by small, computable
radiative corrections, has been
experimentally proven. This relation means that the effective Higgs
(be it fundamental or composite) is indeed a weak isospin doublet.
The Higgs particle has not been found but, in the SM, its mass can well
be larger than the present direct lower limit $m_H \gappeq 114$~GeV (at $95\%$ c.l.)
obtained from  searches at LEP-2.  As we shall see, the radiative corrections
computed in the SM when compared to the data on precision electroweak
tests lead to a clear indication for a light Higgs, not too far from
the present lower bound. The exact experimental upper limit for $m_H$ in the SM depends on the value of the top quark mass $m_t$ (the one-loop radiative corrections are quadratic in $m_t$ and logarithmic in $m_H$).  The CDF and D0 combined value after Run II is at present \cite{ICHEP'06} $m_t= 171.4\pm2.1~GeV$ (it went  down with respect to the value $m_t= 178\pm4.3~GeV$ from Run I and also the experimental error is now sizably reduced). As a consequence the present limit on $m_H$ is more stringent \cite{ewwg}: $m_H < 199~GeV$ (at $95\%$ c.l., after including the information from the 114 GeV direct bound). On the Higgs the LHC will address the following questions : one doublet, more doublets, additional singlets? SM Higgs or SUSY Higgses? Fundamental or composite (of fermions, of WW...)? Pseudo-Goldstone boson of an enlarged symmetry? A manifestation of large extra dimensions (5th component of a gauge boson, an effect of orbifolding or of boundary conditions...)? Or some combination of the above or something so far unthought of?

\section{THEORETICAL BOUNDS ON THE SM HIGGS}

It is well known \cite{zzi}, \cite{zzii}, \cite{aaiiii} that in the SM with only one Higgs doublet a lower limit on
$m_H$ can be derived from the requirement of vacuum stability (or, in milder form, from a moderate instability, compatible with the lifetime of the Universe  \cite{isi}). The limit is a function of $m_t$ and of the energy scale
$\Lambda$ where the model breaks down and new physics appears. The Higgs mass enters because it fixes the initial value of the quartic Higgs coupling $\lambda$ for its running up to the large scale $\Lambda$. Similarly an upper bound on $m_H$ (with mild dependence
on $m_t$) is obtained \cite{eeiiii} from the requirement that for $\lambda$, up to the scale $\Lambda$, no Landau pole appears, or in more explicit terms, that the perturbative description of the theory remains valid. We now briefly recall the derivation of these limits.
	
	The possible instability of the Higgs potential $V[\phi]$ is generated by the quantum loop corrections to the
classical expression of $V[\phi]$. At large $\phi$ the derivative $V'[\phi]$ could become negative and the potential
would become unbound from below. The one-loop corrections to $V[\phi]$ in the SM are well known and change the dominant
term at large $\phi$ according to $\lambda \phi^4 \rightarrow (\lambda +
\gamma~{\rm log}~\phi^2/\Lambda^2)\phi^4$. The one-loop approximation is not enough in this case, because it fails
at large enough $\phi$, when
$\gamma~{\rm log}~\phi^2/\Lambda^2$ becomes of order 1. The renormalization group improved version of the corrected
potential leads to the replacement $\lambda\phi^4 \rightarrow
\lambda(\Lambda)\phi'^4(\Lambda)$ where $\lambda(\Lambda)$ is the running coupling and
$\phi'(\mu) =\phi {\rm exp}\int^t \gamma(t')dt'$, with $\gamma(t)$ being an anomalous dimension function and $t = {\rm
log}\Lambda/v$ ($v$ is the vacuum expectation value
$v = (2\sqrt 2 G_F)^{-1/2}$). As a result, the positivity condition for the potential amounts to the requirement that
the running coupling $\lambda(\Lambda)$ never becomes negative. A more precise calculation, which also takes into
account the quadratic term in the potential, confirms that the requirements of positive
$\lambda(\Lambda)$ leads to the correct bound down to scales $\Lambda$ as low as $\sim$~1~TeV. The running of
$\lambda(\Lambda)$ at one loop is given by: 
\begin{equation}
\frac{d\lambda}{dt} = \frac{3}{4\pi^2} [ \lambda^2 + 3\lambda h^2_t - 9h^4_t + {\rm small~gauge~and~Yukawa~terms }]~,
\label{131}
\end{equation} with the normalization such that at $t=0, \lambda = \lambda_0 = m^2_H/2v^2$ and the top Yukawa coupling
$h_t^0 = m_t/v$. We see that, for $m_H$ small and $m_t$ fixed at its measured value,
$\lambda$ decreases with $t$ and can become negative.  If one requires that
$\lambda$ remains positive up to $\Lambda = 10^{15}$--$10^{19}$~GeV, then the resulting bound on $m_H$ in the SM with
only one Higgs doublet is given by \cite{aaiiii}:
\begin{equation} m_H(\rm{GeV}) > 129.5 + 2.1 \left[ m_t - 171.4 \right] - 4.5~\frac{\alpha_s(m_Z) - 0.118}{0.006}~.
\label{25h}
\end{equation}
Note that this limit is evaded in models with more Higgs doublets. In this case the limit applies to some average mass but the lightest Higgs particle can well be below, as it is the case in the minimal SUSY extension of the SM (MSSM).

The upper limit on
the Higgs mass in the SM is clearly important for assessing the chances of success of the LHC as an accelerator designed
to solve the Higgs problem. The upper limit \cite{eeiiii} arises from the requirement that the Landau pole associated with the non
asymptotically free behaviour of the $\lambda \phi^4$ theory does not occur below the scale $\Lambda$. The initial value of
$\lambda$ at the weak scale increases with $m_H$ and the derivative is positive at large $\lambda$ (because of the positive $\lambda^2$ term - the $\lambda \varphi^4$ theory is not asymptotically free - overwhelms the negative top-Yukawa term). Thus if $m_H$ is too large the point where $\lambda$ computed from the perturbative beta function becomes infinite (the Landau pole) occurs at too low an energy.  Of course in the vicinity of the Landau pole the 2-loop evaluation of the beta function is not reliable. Indeed the limit indicates the frontier of the domain where the theory is well described by the perturbative  expansion. Thus the quantitative evaluation of the limit is only indicative, although it has been to some extent supported by simulations of the Higgs sector of the EW theory on the lattice. For the upper limit on $m_H$ one finds \cite{eeiiii}
$m_H\lappeq 180~GeV$ for $\Lambda\sim M_{GUT}-M_{Pl}$ and $m_H\lappeq 0.5-0.8~TeV$ for $\Lambda\sim
1~TeV$. Actually, for
$m_t \sim$ 171~GeV, only a small range of values for $m_H$ is allowed, $130 < m_H <~\sim 200$~GeV, if the SM holds up
to $\Lambda \sim M_{GUT}$ or $M_{Pl}$. This upper limit implies that the SM Higgs cannot escape detection at the LHC.

\section{PRECISION TESTS OF THE STANDARD ELECTROWEAK THEORY}

\begin{figure}
\centering
\includegraphics[width=85mm]{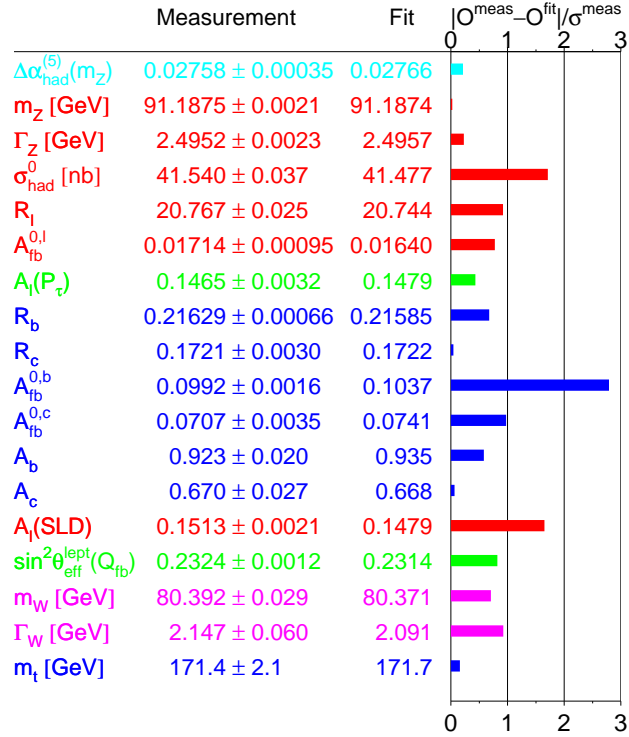}
\caption{Pulls, Summer'06} \label{pull}
\end{figure}

The results of the electroweak precision tests as well as of the searches
for the Higgs boson and for new particles performed at LEP and SLC are now available in  final form \cite{ewwg} \cite{AG}.  Taken together
with the measurements of $m_t$, $m_W$ and the searches for new physics at the Tevatron, and with some other data from low
energy experiments, they form a very stringent set of precise constraints to be compared with the SM or with
any of its conceivable extensions. 
All high energy precision tests of the SM are summarized in fig.~\ref{pull} \cite{ewwg}.
For the analysis of electroweak data in the SM one starts from the
input parameters: as in any renormalizable theory masses and couplings
have to be specified from outside. One can trade one parameter for
another and this freedom is used to select the best measured ones as
input parameters. Some of them, $\alpha$, $G_F$ and
$m_Z$, are very precisely known, some other ones, $m_{f_{light}}$,
$m_t$ and $\alpha_s(m_Z)$ are far less well determined while $m_H$ is
largely unknown.  Among the light fermions, the quark masses are badly known, but
fortunately, for the calculation of radiative corrections, they can be
replaced by $\alpha(m_Z)$, the value of the QED running coupling at
the Z mass scale. The value of the hadronic contribution to the
running, $\dalhad$, reported in Fig.~\ref{pull}, is obtained
through dispersion relations from the data on $e^+e^-\rightarrow \rm{hadrons}$ at
low centre-of-mass energies~\cite{ewwg}. From the input parameters
one computes the radiative corrections to a sufficient precision to
match the experimental accuracy. Then one compares the theoretical
predictions with the data for the numerous observables which have been
measured, checks the consistency of the theory and derives constraints
on $m_t$, $\alpha_s(m_Z)$ and $m_H$.

The computed radiative corrections include the complete set of
one-loop diagrams, plus some selected large subsets of two-loop
diagrams and some sequences of resummed large terms of all orders
(large logarithms and Dyson resummations). In particular large
logarithms, e.g., terms of the form $(\alpha/\pi ~{\rm
ln}~(m_Z/m_{f_\ell}))^n$ where $f_{\ell}$ is a light fermion, are
resummed by well-known and consolidated techniques based on the
renormalisation group. For example, large logarithms dominate the
running of $\alpha$ from $m_e$, the electron mass, up to $m_Z$, which
is a $6\%$ effect, much larger than the few per mil contributions of
purely weak loops.  Also, large logs from initial state radiation
dramatically distort the line shape of the Z resonance observed at
LEP-1 and SLC and have been accurately taken into account in the
measurement of the Z mass and total width.

Among the one loop EW radiative corrections a remarkable class of
contributions are those terms that increase quadratically with the top
mass.  The large sensitivity of radiative corrections to $m_t$ arises
from the existence of these terms. The quadratic dependence on $m_t$
(and possibly on other widely broken isospin multiplets from new
physics) arises because, in spontaneously broken gauge theories, heavy
loops do not decouple. On the contrary, in QED or QCD, the running of
$\alpha$ and $\alpha_s$ at a scale $Q$ is not affected by heavy quarks
with mass $M \gg Q$. According to an intuitive decoupling
theorem~\cite{AC}, diagrams with heavy virtual particles of mass
$M$ can be ignored for $Q \ll M$ provided that the couplings do not
grow with $M$ and that the theory with no heavy particles is still
renormalizable. In the spontaneously broken EW gauge theories both
requirements are violated. First, one important difference with
respect to unbroken gauge theories is in the longitudinal modes of
weak gauge bosons. These modes are generated by the Higgs mechanism,
and their couplings grow with masses (as is also the case for the
physical Higgs couplings). Second, the theory without the top quark is
no more renormalizable because the gauge symmetry is broken if the b
quark is left with no partner (while its couplings show that the weak
isospin is 1/2). Because of non decoupling precision tests of the
electroweak theory may be sensitive to new physics even if the new
particles are too heavy for their direct production.

While radiative corrections are quite sensitive to the top mass, they
are unfortunately much less dependent on the Higgs mass. If they were
sufficiently sensitive, by now we would precisely know the mass of the
Higgs. However, the dependence of one loop diagrams on $m_H$ is only
logarithmic: $\sim G_F m_W^2\log(m_H^2/m_W^2)$. Quadratic terms $\sim
G_F^2 m_H^2$ only appear at two loops and are too small to be
important. The difference with the top case is that $m_t^2-m_b^2$ is a
direct breaking of the gauge symmetry that already affects the
relevant one loop diagrams, while the Higgs couplings to gauge bosons
are "custodial-SU(2)" symmetric in lowest order.

\begin{figure}
\centering
\includegraphics[width=85mm]{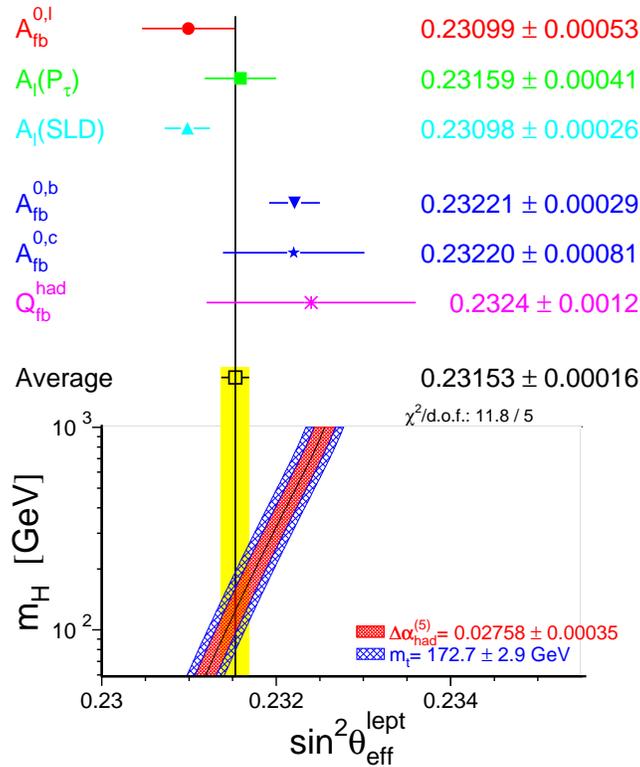}
\caption{Effective electroweak
mixing angle $\swsqeffl$ derived from measurement results depending on
lepton couplings only (above) and also quark couplings (below).  } \label{sef2}
\end{figure}

The various asymmetries determine the effective electroweak mixing
angle for leptons with highest sensitivity.  The results on
$\swsqeffl$ are compared in Figure~\ref{sef2}. The weighted
average of these six results, including small correlations, is:
\begin{eqnarray}
\swsqeffl & = & 0.23153\pm0.00016 \,.
\label{eq:sin2teff}
\end{eqnarray}
Note, however, that this average has a $\chi^2$ of 11.8 for 5 degrees
of freedom, corresponding to a probability of 3.7\%. The $\chi^2$ is
pushed up by the two most precise measurements of $\swsqeffl$, namely
those derived from the measurements of $A_l$ by SLD, dominated by the
left-right asymmetry $A_{LR}$, and of the forward-backward asymmetry
measured in $b \bar b$ production at LEP, $A^b_{FB}$, which differ by about
3.2 $\sigma$'s.   In
general, there appears to be a discrepancy between $\swsqeffl$
measured from leptonic asymmetries ($(\sin^2\theta_{\rm eff})_l$) and
from hadronic asymmetries ($(\sin^2\theta_{\rm eff})_h$), as seen from Figure~\ref{sef2}. In fact, the result from $A_{LR}$ is in good
agreement with the leptonic asymmetries measured at LEP, while all
hadronic asymmetries, though their errors are large, are better
compatible with the result of $A^b_{FB}$.
This very unfortunate fact makes the interpretation of precision tests less sharp and some perplexity remains: is it an experimental error or a signal of some new physics?
\begin{figure}
\centering
\includegraphics[width=85mm]{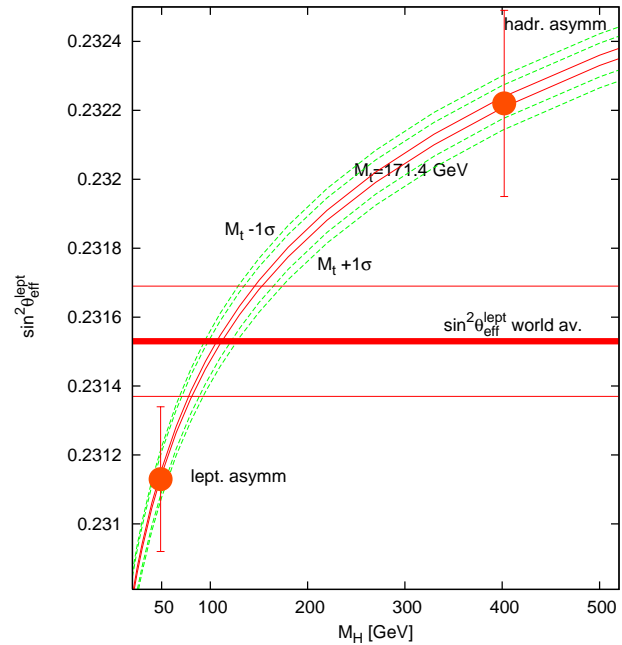}
\caption{The data for $\sin^2\theta_{\rm eff}^{\rm lept}$ are
plotted vs $m_H$. For presentation purposes the measured points are
shown each at the $m_H$ value that would ideally correspond to it
given the central value of $m_t$ (updated from \cite{P-Gambino}). } \label{s2mH}
\end{figure}
The situation is shown in Figure~\ref{s2mH}~\cite{P-Gambino}.
The values of $(\sin^2\theta_{\rm eff})_l$, $(\sin^2\theta_{\rm
eff})_h$ and their formal combination are shown each at the $m_H$
value that would correspond to it given the central value of $m_t$.
Of course, the value for $m_H$ indicated by each $\swsqeffl$ has an
horizontal ambiguity determined by the measurement error and the width
of the $\pm1\sigma$ band for $m_t$.  Even taking this spread into
account it is clear that the implications on $m_H$ are sizably
different.  

One might imagine that some new physics effect could be
hidden in the $\mathrm{Z b \bar b}$ vertex.  Like for the top quark
mass there could be other non decoupling effects from new heavy states
or a mixing of the b quark with some other heavy quark.  However, it
is well known that this discrepancy is not easily explained in terms
of some new physics effect in the $\mathrm{Z b \bar b}$ vertex. In
fact, $A^b_{FB}$ is the product of lepton- and b-asymmetry factors:
$A^b_{FB}=(3/4){\cal A}_e {\cal A}_b$.  The sensitivity of $A^b_{FB}$ to ${\cal A}_b$ is
limited, because the ${\cal A}_e$ factor is small, so that a rather large
change of the b-quark couplings with respect to the SM is needed in
order to reproduce the measured discrepancy (precisely a $\sim 30\%$
change in the right-handed coupling $g^b_R$, an effect too large to be a loop
effect but which could be produced at the tree level, e.g., by mixing
of the b quark with a new heavy vectorlike quark \cite{CTW}).  But
 this effect is not confirmed by the direct measurement
of ${\cal A}_b$ performed at SLD using the left-right polarized b asymmetry,
even taking into account the moderate precision of this result. No deviation is manifest in the accurate measurement of $R_b \propto
g_{\mathrm{Rb}}^2+g_{\mathrm{Lb}}^2$.   Thus, even introducing an
ad hoc mixing the overall fit of $A^b_{FB}$, ${\cal A}_b$ and $R_b$ is not terribly good, but we cannot
exclude this possibility completely.  Alternatively, the observed
discrepancy could be due to a large statistical fluctuation or an
unknown experimental problem. In any case the effective ambiguity in the measured value of
$\swsqeffl$ is actually larger than the nominal error, reported in
Eq.~\ref{eq:sin2teff}, obtained from averaging all the existing
determinations.

We now discuss fitting the data in the SM. One can think of different
types of fit, depending on which experimental results are included or
which answers one wants to obtain. For example, in
Table~\ref{tab:fit:result} we present in column~1 a fit of all Z pole
data plus $m_W$  (this is interesting as it shows the value
of $m_t$ obtained indirectly from radiative corrections, to be
compared with the value of $m_t$ measured in production experiments),
in column~2 a fit of all Z pole data plus $m_t$ (here it is $m_W$
which is indirectly determined), and, finally, in column~3 a fit of
all the data listed in Fig. 1 (which is the most
relevant fit for constraining $m_H$).  From the fit in column~1 of
Table~\ref{tab:fit:result} we see that the extracted value of $m_t$ is
in good agreement with the direct measurement (see the value reported in
Fig. 1).  Similarly we see that the direct
determination of $m_W$ reported in Fig. 1 is still a bit larger with respect to the value from the fit in
column~2 (although the direct value of $m_W$ went down recently).  We have seen that quantum corrections depend only
logarithmically on $m_H$.  In spite of this small sensitivity, the
measurements are precise enough that one still obtains a quantitative
indication of the mass range. From the fit in column~3 we obtain:
$\log_{10}{m_H(\GeV)}=1.93\pm 0.17$ (or $m_H=85^{+39}_{-28}~\GeV$). We see that the central value of $m_H$ from the fit is below the lower limit on the SM Higgs mass from direct searches $m_H\gappeq 114~\GeV$, but within $1\sigma$ from this bound. If we had reasons to remove the result on $A^b_{FB}$  from the fit, the fitted value of $m_H$ would move down: $m_H=55^{+31}_{-21}~\GeV$ \cite{gru}, further away from the lower limit.
\begin{table}[tb]
\begin{center}
\renewcommand{\arraystretch}{1.3}
\begin{tabular}{|l||c|c|c|}
\hline 
Fit       & 1 & 2 & 3 \\
\hline
\hline
Measurements      &$m_W$         &$m_t$            &$m_t,~m_W$\\
\hline
\hline
$m_t~(\GeV)$      &$177.6^{+12}_{-9}$&$171.4\pm2.1$    &$171.7\pm2.0$\\
$m_H~(\GeV)$      &$137^{+228}_{-76}$    &$103^{+54}_{-37}$&$85^{+39}_{-28}$\\
$\log~[m_H(\GeV)]$&$2.14\pm{+0.39}$&$2.01\pm0.19$    &$1.93\pm0.17$ \\
$\alpha_s(m_Z)$   &$0.1190\pm0.0028$     &$0.1190\pm0.0027$&$0.1186\pm0.0026$\\
\hline
$m_W~(\MeV)$      &$80380 \pm 21$    &$80361 \pm 20$   &$80371 \pm 16$  \\
\hline
\end{tabular}
\caption[]{ Standard Model fits of electroweak data. All fits use the
Z pole results and $\dalhad$ as listed in Fig.~\ref{pull}. In
addition, the measurements listed on top of each column are included as
well. The fitted W mass is also shown \cite{ewwg} (the directly measured value is
$m_W=80392 \pm 29 \MeV $).}
\label{tab:fit:result}
\end{center}
\end{table} 

\begin{figure}
\centering
\includegraphics[width=85mm]{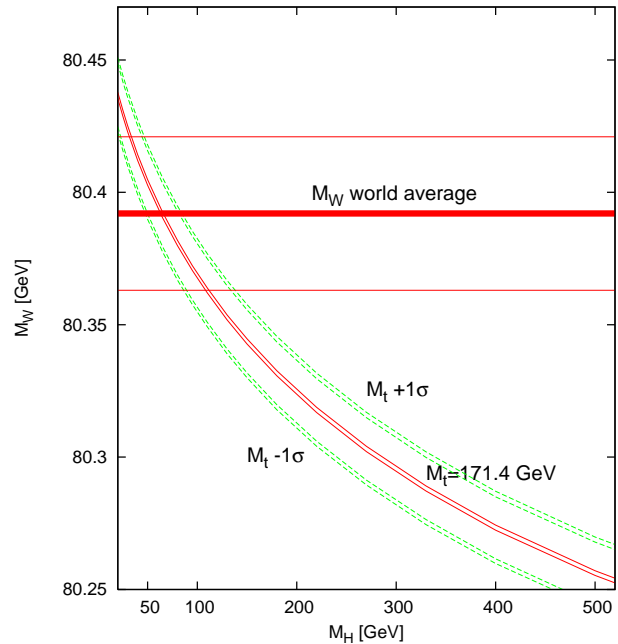}
\caption{The world average for $m_W$ is
plotted vs $m_H$ (updated from \cite{P-Gambino}). } \label{wmH}
\end{figure}

We have already observed that the experimental value of $m_W$ (with
good agreement between LEP and the Tevatron) is a bit high compared to
the SM prediction (see Figure~\ref{wmH}). The value of $m_H$
indicated by $m_W$ is on the low side, just in the same interval as
for $\sin^2\theta_{\rm eff}^{\rm lept}$ measured from leptonic
asymmetries.  The recent decrease of the experimental value of $m_t$
maintains the tension between the experimental
values of $m_W$ and $\sin^2\theta_{\rm eff}^{\rm lept}$ measured from
leptonic asymmetries on one side and the lower limit on $m_H$ from
direct searches on the other side~\cite{cha}, \cite {acggr}.  
 
With all these words of caution in mind it remains true that on the whole the SM performs
rather well, so that it is fair to say that no clear indication for
new physics emerges from the data. Actually the result of precision tests on the Higgs mass is particularly remarkable. The value of
$\log_{10}{m_H(\GeV)}$ is, within errors, within the small window between
$\sim 2$ and $\sim 3$ which is allowed, on the one side, by the direct
search limit ($m_H\gappeq 114~\GeV$ from LEP-2~\cite{ewwg}),
and, on the other side, by the theoretical upper limit on the Higgs
mass in the minimal SM  \cite{eeiiii}, $m_H\lappeq 600-800~\GeV$.

Thus the whole picture of a perturbative theory with a fundamental
Higgs is well supported by the data on radiative corrections. It is
important that there is a clear indication for a particularly light
Higgs: at $95\%$ c.l. $m_H\lappeq 199~\GeV$.  This is quite
encouraging for the ongoing search for the Higgs particle.  More in
general, if the Higgs couplings are removed from the Lagrangian the
resulting theory is non renormalizable. A cutoff $\Lambda$ must be
introduced. In the quantum corrections $\log{m_H}$ is then replaced by
$\log{\Lambda}$ plus a constant. The precise determination of the
associated finite terms would be lost (that is, the value of the mass
in the denominator in the argument of the logarithm).  A heavy Higgs
would need some conspiracy: the finite terms, different in
the new theory from those of the SM, should accidentally compensate
for the heavy Higgs in a few key parameters of the radiative
corrections (mainly $\epsilon_1$ and $\epsilon_3$, see, for example,
\cite{eps}).  Alternatively, additional new physics, for example in
the form of effective contact terms added to the minimal SM
lagrangian, should accidentally do the compensation, which again needs
some sort of conspiracy, although this possibility is not so unlikely to be apriori discarded.

\section{THE PHYSICS OF FLAVOUR}

\begin{figure}
\centering
\includegraphics[width=100mm]{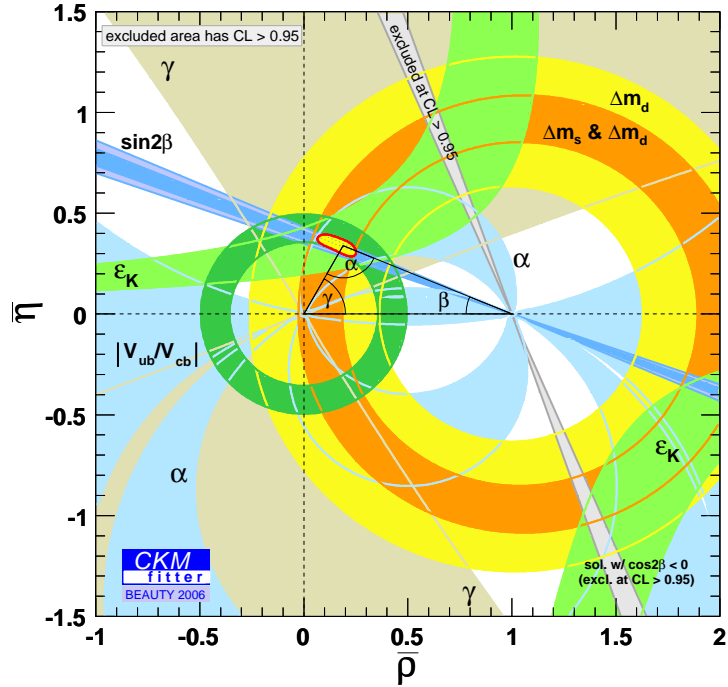}
\caption{Constraints in the $\bar \rho,\bar \eta$ plane including  the most recent $\alpha$, $\gamma$ and $\Delta M_s$ inputs in the global CKM fit  \cite{CKMfitter}). } \label{CKM}
\end{figure}

In the last decade great progress in different areas of flavour physics has also been achieved. In the quark sector, the important results of a generation of frontier experiments have become available, obtained at B factories and at accelerators. The hope of these experiments was to detect departures from the CKM picture of mixing and of CP violation as  signals of new physics. But so far the observed B mixing and CP violation agree very well with the SM predictions based on the CKM matrix \cite{fle}. The recent measurement of $\Delta m_s$ by CDF and D0, in fair agreement with the SM expectation, has closed another door for new physics. In quantitative terms all measurements are in agreement with the CKM description of mixing and CP violation as shown by Fig. \ref{CKM}.  It is only in channels that are forbidden at tree level and occur through penguin loops (as is the case for $B \rightarrow  \pi K$ modes) that some deviation could still be hidden. The amazing performance of the SM in flavour changing transitions and for CP violation in K and B decays poses a strong constraint on all proposed models of new physics. 

In the leptonic sector the study of neutrino oscillations has led to the discovery that at least two neutrinos are not massless and to the determination of the mixing matrix \cite{alfe}. Neutrinos are not all massless but their masses are very small. Probably masses are small because $\nu$Õs are Majorana particles, and, by the see-saw mechanism, their masses are inversely proportional to the large scale $M$ where lepton number ($L$) violation occurs (as expected in GUT's). Indeed the value of $M\sim m_{\nu R}$ from experiment is compatible with being close to $M_{GUT} \sim 10^{14}-10^{15}~GeV$, so that neutrino masses fit well in the GUT picture and actually support it.  It was realized that decays of heavy $\nu_R$ with CP and L violation can produce a B-L asymmetry. The range of neutrino masses indicated by neutrino phenomenology turns out to be perfectly compatible with the idea of baryogenesis via leptogenesis \cite{buch}. This elegant model for baryogenesis has by now replaced the idea of baryogenesis near the weak scale, which has been strongly disfavoured by LEP. It is remarkable that we now know the neutrino mixing matrix with good accuracy. Two mixing angles are large and one is small. The atmospheric angle $\theta_{23}$ is large, actually compatible with maximal but not necessarily so: at $3\sigma$: $0.31 \leq \sin^2{\theta_{23}}\leq 0.72$ with central value around  $0.5$. The solar angle $\theta_{12}$ is large, $\sin^2{\theta_{12}}\sim 0.3$, but certainly not maximal (by more than 5$\sigma$). The third angle $\theta_{13}$, strongly limited mainly by the CHOOZ experiment, has at present a $3\sigma$ upper limit given by about $\sin^2{\theta_{13}}\leq 0.08$. While these discoveries are truly remarkable, it is somewhat depressing that the detailed knowledge of both the quark and the neutrino mixings has not led so far to a compelling solution of the dynamics of fermion masses and mixings: our models can reproduce, actually along different paths, the observed values, but we do not really understand their mysterious pattern.

The only plausible indication of new physics could be the apparent discrepancy from the SM prediction of the muon (g-2) measurement by the  BNL experiment \cite{BNL}. The deviation is reported at the level of  $3.3 \sigma$'s \cite{eid}. I said that the anomaly is plausible. In fact this observable is sufficiently sensitive, so that it is reasonable that the detection of new physics could occur here. There are reasonable models that could accomodate the observed discrepancy, for example SUSY with light electroweak s-particles and moderately large $\tan{\beta}$ \cite {acggr}. There are however some words of caution. The dominant error is the theoretical error associated with the hadronic contributions to the vacuum polarization and to the light by light scattering 4-point function \cite{pas}. As usual for theoretical errors, they are to some extent debatable and authors tend to emphasize the importance of each progress, e.g. in the data on hadronic cross sections, by decreasing the theoretical error more and more. Also there is a difference between the predictions on vacuum polarization based on $e^+e^-$ crosssections versus hadronic $\tau$ decays (the g-2 discrepancy is much less if estimated from the $\tau$ data). This difference is not understood. 

\section{PROBLEMS OF THE STANDARD MODEL}

No signal of new physics has been
found neither in electroweak precision tests nor in flavour physics. Given the success of the SM why are we not satisfied with this beautiful theory? Why not just find the Higgs particle,
for completeness, and declare that particle physics is closed? The reason is that there are
both conceptual problems and phenomenological indications for physics beyond the SM. On the conceptual side the most
obvious problems are that quantum gravity is not included in the SM and the related hierarchy problem. Among the main
phenomenological hints for new physics we can list coupling unification, dark matter, neutrino masses, 
baryogenesis and the cosmological vacuum energy.

The computed evolution with energy
of the effective SM gauge couplings clearly points towards the unification of the electro-weak and strong forces (Grand
Unified Theories: GUT's) at scales of energy
$M_{GUT}\sim  10^{15}-10^{16}~ GeV$ which are close to the scale of quantum gravity, $M_{Pl}\sim 10^{19}~ GeV$.  One is led to
imagine  a unified theory of all interactions also including gravity (at present superstrings provide the best attempt at such
a theory). Thus GUT's and the realm of quantum gravity set a very distant energy horizon that modern particle theory cannot
ignore. Can the SM without new physics be valid up to such large energies? One can imagine that some obvious problems could be postponed to the more fundamental theory at the Planck mass. For example, the explanation of the three generations of fermions and the understanding of fermion masses and mixing angles can be postponed. But other problems must find their solution in the low energy theory. In particular, the structure of the
SM could not naturally explain the relative smallness of the weak scale of mass, set by the Higgs mechanism at $\mu\sim
1/\sqrt{G_F}\sim  250~ GeV$  with $G_F$ being the Fermi coupling constant. This so-called hierarchy problem is due to the instability of the SM with respect to quantum corrections. This is related to
the
presence of fundamental scalar fields in the theory with quadratic mass divergences and no protective extra symmetry at
$\mu=0$. For fermion masses, first, the divergences are logarithmic and, second, they are forbidden by the $SU(2)\bigotimes
U(1)$ gauge symmetry plus the fact that at
$m=0$ an additional symmetry, i.e. chiral  symmetry, is restored. Here, when talking of divergences, we are not
worried of actual infinities. The theory is renormalizable and finite once the dependence on the cut off $\Lambda$ is
absorbed in a redefinition of masses and couplings. Rather the hierarchy problem is one of naturalness. We can look at the
cut off as a parameterization of our ignorance on the new physics that will modify the theory at large energy
scales. Then it is relevant to look at the dependence of physical quantities on the cut off and to demand that no
unexplained enormously accurate cancellations arise. 

The hierarchy problem can be put in very practical terms (the "little hierarchy problem"): loop corrections to the Higgs mass squared are
quadratic in $\Lambda$. The most pressing problem is from the top loop.
 With $m_h^2=m^2_{bare}+\delta m_h^2$ the top loop gives 
 \begin{eqnarray}
\delta m_{h|top}^2\sim -\frac{3G_F}{2\sqrt{2} \pi^2} m_t^2 \Lambda^2\sim -(0.2\Lambda)^2 \label{top}
\end{eqnarray}

If we demand that the correction does not exceed the light Higgs mass indicated by the precision tests, $\Lambda$ must be
close, $\Lambda\sim o(1~TeV)$. Similar constraints arise from the quadratic $\Lambda$ dependence of loops with gauge bosons and
scalars, which, however, lead to less pressing bounds. So the hierarchy problem demands new physics to be very close (in
particular the mechanism that quenches the top loop). Actually, this new physics must be rather special, because it must be
very close, yet its effects are not clearly visible (the "LEP Paradox" \cite{BS}). 

\section{AVENUES BEYOND THE STANDARD MODEL}

Examples of proposed classes of solutions
for the hierarchy problem are:

¥ $\bf{Supersymmetry.}$ In the limit of exact boson-fermion symmetry the quadratic divergences of bosons cancel so that
only log divergences remain. However, exact SUSY is clearly unrealistic. For approximate SUSY (with soft breaking terms),
which is the basis for all practical models, $\Lambda$ is replaced by the splitting of SUSY multiplets, $\Lambda\sim
m_{SUSY}-m_{ord}$. In particular, the top loop is quenched by partial cancellation with s-top exchange, so the s-top cannot be too heavy.

¥ $\bf{Technicolor.}$ The Higgs system is a condensate of new fermions. There is no fundamental scalar Higgs sector, hence no
quadratic divergences associated to the $\mu^2$ mass in the scalar potential. This mechanism needs a very strong binding force,
$\Lambda_{TC}\sim 10^3~\Lambda_{QCD}$. It is  difficult to arrange that such nearby strong force is not showing up in
precision tests. Hence this class of models has been disfavoured by LEP, although some special class of models have been devised aposteriori, like walking TC, top-color assisted TC etc (for recent reviews, see, for example, \cite{L-C}). 

$\bf{Large~extra~dimensions.}$ The idea is that $M_{Pl}$ appears very large, or equivalently that gravity appears very weak,
because we are fooled by hidden extra dimensions so that the real gravity scale is reduced down to a lower scale, even possibly down to
$o(1~TeV)$. This possibility is very exciting in itself and it is really remarkable that it is compatible with experiment.

¥ $\bf{"Little~Higgs" models.}$ In these models the Higgs is a pseudo-Goldstone boson and extra symmetries allow $m_h\not= 0$ only at two-loop level, so that $\Lambda$
can be as large as
$o(10~TeV)$ with the Higgs within present bounds (the top loop is quenched by exchange of heavy vectorlike new  quarks with charge 2/3).

In the following we briefly comment in turn on these possibilities.

\section{SUPERSYMMETRY}

SUSY models are the most developed and most widely accepted. Many theorists consider SUSY as established at the Planck
scale $M_{Pl}$. So why not to use it also at low energy to fix the hierarchy problem, if at all possible? It is interesting
that viable models exist. The necessary SUSY breaking can be introduced through soft
terms that do not spoil the good convergence properties of the theory. Precisely those terms arise from
supergravity when it is spontaneoulsly broken in a hidden sector. This is the case of the MSSM \cite{Martin} with minimal particle content. Of course, minimality is only a simplicity assumption that could possibly be relaxed. For example, adding an additional Higgs singlet S considerably helps in addressing naturalness constraints \cite{nmssm}, \cite{barbie}. Minimal versions or, even more, versions with additional constraints like the constrained MSSM (CMSSM) (where simple conditions at the GUT scale are in addition assumed) are economic in terms of new parameters but could be to some extent misleading. Still, the MSSM 
is a completely specified,
consistent and computable theory which is compatible with all precision electroweak tests. In this
most traditional approach SUSY is broken in a hidden sector and the scale of SUSY breaking is very
large of order
$\Lambda\sim\sqrt{G^{-1/2}_F M_{Pl}}$. But since the hidden sector only communicates with the visible sector
through gravitational interactions the splitting of the SUSY multiplets is much smaller, in the TeV
energy domain, and the Goldstino is practically decoupled. 
But alternative mechanisms of SUSY breaking are also being considered. In one alternative scenario \cite{gau} the (not so
much) hidden sector is connected to the visible one by ordinary gauge interactions. As these are much
stronger than the gravitational interactions, $\Lambda$ can be much smaller, as low as 10-100
TeV. It follows that the Goldstino is very light in these models (with mass of order or below 1 eV
typically) and is the lightest, stable SUSY particle, but its couplings are observably large. The radiative
decay of the lightest neutralino into the Goldstino leads to detectable photons. The signature of photons comes
out naturally in this SUSY breaking pattern: with respect to the MSSM, in the gauge mediated model there are typically
more photons and less missing energy. The main appeal of gauge mediated models is a better protection against
flavour changing neutral currents but naturality problems tend to increase. As another possibility it has been
pointed out that there are pure gravity contributions to soft masses that arise from anomalies in gravity theory  \cite{ano}. In the assumption that these terms are dominant the associated spectrum and phenomenology have been
studied. In this case gaugino masses are proportional to gauge coupling beta functions, so that the gluino is much heavier
than the electroweak gauginos. 

What is really unique to SUSY with respect to all other extensions of the SM listed above is that the MSSM or
other non minimal SUSY models are well defined and computable up to $M_{Pl}$ and, moreover, are not only compatible but actually 
quantitatively supported by coupling unification and GUT's. At present the most direct
phenomenological evidence in favour of supersymmetry is obtained from the unification of couplings in GUTs.
Precise LEP data on $\alpha_s(m_Z)$ and $\sin^2{\theta_W}$ show that
standard one-scale GUTs fail in predicting $\alpha_s(m_Z)$ given $\sin^2{\theta_W}$ 
and $\alpha(m_Z)$ while SUSY GUTs are compatible with the present, very precise,
experimental results (of course, the ambiguities in the prediction are larger because of our ignorance of the SUSY spectrum). If one starts from the known values of
$\sin^2{\theta_W}$ and $\alpha(m_Z)$, one finds \cite{LP} for $\alpha_s(m_Z)$ the results:
$\alpha_s(m_Z) = 0.073\pm 0.002$ for Standard GUTs and $\alpha_s(m_Z) = 0.129\pm0.010$ for SUSY~ GUTs
to be compared with the world average experimental value $\alpha_s(m_Z) =0.118\pm0.002$. Another great asset of SUSY GUT's
is that proton decay is much slowed down with respect to the non SUSY case. First, the unification mass $M_{GUT}\sim~\rm{few}~
10^{16}~GeV$, in typical SUSY GUT's, is about 20-30 times larger than for ordinary GUT's. This makes p decay via gauge
boson exchange negligible and the main decay amplitude arises from dim-5 operators with higgsino exchange, leading to a
rate close but still compatible with existing bounds (see, for example,\cite{AFM}). It is also important that SUSY provides an excellent dark matter candidate, the neutralino.  We finally recall that the range of neutrino masses as indicated by oscillation experiments, when interpreted in the see-saw mechanism, point to $M_{GUT}$ and give additional support to GUTs \cite{alfe}.

In spite of all these virtues it is true that the lack of SUSY signals at LEP and the lower limit on $m_H$ pose problems
for the MSSM. The lightest Higgs particle is predicted in the MSSM to be below $m_h\lappeq~130~GeV$ (with the esperimental value of $m_t$ going down the upper limit is slightly decreased). The limit on the SM
Higgs
$m_H\gappeq~114~GeV$ considerably restricts the available parameter space of the MSSM requiring relatively large $\tan\beta$
($\tan\beta\gappeq~2-3$: at tree level $m^2_h=m^2_Z\cos^2{2\beta}$) and rather heavy s-top (the loop corrections
increase with $m_t^4 \log{\tilde{m_t^2}}$). But we have seen that a heavy s-top is unnatural, because it enters quadratically in the radiative corrections to $\delta m_{h|top}^2$. Stringent naturality constraints also follow from imposing that the electroweak
symmetry breaking occurs at the right energy scale: in SUSY models the breaking is induced by the running of the $H_u$ mass
starting from a common scalar mass $m_0$ at $M_{GUT}$. The squared Z mass $m_Z^2$ can be expressed as a linear
combination of the SUSY parameters $m_0^2$, $m_{1/2}^2$, $A^2_t$, $\mu^2$,... with known coefficients. Barring
cancellations that need fine tuning, the SUSY parameters, hence the SUSY s-partners cannot be too heavy. The LEP limits,
in particular the chargino lower bound $m_{\chi+}\gappeq~100~GeV$, are sufficient to eliminate an important region of the
parameter space, depending on the amount of allowed fine tuning. For example, models based on gaugino universality at the
GUT scale, like the CMSSM, are discarded unless a fine tuning by at least a factor of ~20 is not allowed. Without gaugino
universality \cite{kane} the strongest limit remains on the gluino mass: $m_Z^2\sim 0.7~m_{gluino}^2+\dots$ which is still
compatible with the present limit $m_{gluino}\gappeq~240~GeV$ from the TeVatron.

\section{LARGE EXTRA DIMENSIONS}

The non discovery of SUSY at LEP has given further impulse to the quest for new ideas on physics beyond the SM. Large extra
dimensions \cite{Jo} models are  among the most interesting new directions in model building. Large
extra dimension models propose to solve the hierarchy problem by bringing gravity down from $M_{Pl}$ to $m\sim~o(1~TeV)$ where
$m$ is the string scale. Inspired by string theory one assumes that some compactified extra dimensions are sufficiently large
and that the SM fields are confined to a 4-dimensional brane immersed in a d-dimensional bulk while gravity, which feels the
whole geometry, propagates in the bulk. We know that the Planck mass is large because gravity is weak: in fact $G_N\sim
1/M_{Pl}^2$, where
$G_N$ is Newton constant. The idea is that gravity appears so weak because a lot of lines of force escape in extra
dimensions. Assume you have $n=d-4$ extra dimensions with compactification radius $R$. For large distances, $r>>R$, the
ordinary Newton law applies for gravity: in natural units, the force between two units of mass is $F\sim G_N/r^2\sim 1/(M_{Pl}^2r^2)$. At short distances,
$r\lappeq R$, the flow of lines of force in extra dimensions modifies Gauss law and $F^{-1}\sim m^2(mr)^{d-4}r^2$. By
matching the two formulas at $r=R$ one obtains $(M_{Pl}/m)^2=(Rm)^{d-4}$. For $m\sim~1~TeV$ and $n=d-4$ one finds that
$n=1$ is excluded ($R\sim 10^{15} cm$), for $n=2~R$  is at the edge of present bounds $R\sim~1~ mm$, while for $n=4,6$,
$R\sim~10^{-9}, 10^{-12}~cm$.  In all these models a generic feature is the occurrence of Kaluza-Klein (KK) modes.
Compactified dimensions with periodic boundary conditions, as for quantization in a box, imply a discrete spectrum with
momentum $p=n/R$ and mass squared $m^2=n^2/R^2$. There are many versions of these models. The SM brane can itself have a
thickness $r$ with $r<\sim~10^{-17}~cm$ or $1/r>\sim~1~TeV$, because we know that quarks and leptons are pointlike down to
these distances, while for gravity in the bulk there is no experimental counter-evidence down to $R<\sim~0.1~mm$ or
$1/R>\sim~10^{-3}~eV$. In case of a thickness for the SM brane there would be KK recurrences for SM fields, like $W_n$,
$Z_n$ and so on in the TeV region and above. There are models with factorized metric ($ds^2=\eta_{\mu
\nu}dx^{\mu}dx^{\nu}+h_{ij}(y)dy^idy^j$, where y (i,j) denotes the extra dimension coordinates (and indices), or models
with warped metric ($ds^2=e^{-2kR|\phi|} \eta_{\mu \nu}dx^{\mu}dx^{\nu}-R^2\phi^2$ \cite{RS}.
In any case there are the towers of KK recurrences of the graviton. They
are gravitationally coupled but there are a lot of them that sizably couple, so that the net result is a modification
of cross-sections and the presence of missing energy. 

Large extra dimensions provide a very exciting scenario \cite{FeAa}. Already it is remarkable that this possibility is
compatible with experiment. However, there are a number of criticisms that can be brought up. First, the hierarchy problem is
more translated in new terms rather than solved. In fact the basic relation $Rm=(M_{Pl}/m)^{2/n}$ shows that $Rm$, which one
would apriori expect to be $0(1)$, is instead ad hoc related to the large ratio $M_{Pl}/m$.  In this respect  the Randall-Sundrum variety is more appealing because the hierarchy suppression $m_W/M_{Pl}$ could arise from the warping factor $e^{-2kR|\phi|}$, for not too large values of $kR$. The question of whether these values of $kR$ are reasonable has been discussed in ref. \cite{GW}, which offer the best support to the solution of the hierarchy problem in this context. Also it is
not clear how extra dimensions can by themselves solve the LEP paradox (the large top loop corrections should be
controlled by the opening of the new dimensions and the onset of gravity): since
$m_H$ is light
$\Lambda\sim 1/R$ must be relatively close. But precision tests put very strong limits on $\Lambda$. In fact in typical
models of this class there is no mechanism to sufficiently quench the corrections. While no simple, realistic model has
yet emerged as a benchmark, it is attractive to imagine that large extra dimensions could be a part of the truth,
perhaps coupled with some additional symmetry or even SUSY. The Randall-Sundrum warped geometry has become the common framework for many attempts in this direction.

In the  general context of extra dimensions an interesting direction of development is the study of symmetry breaking by orbifolding and/or boundary conditions. These are models where a larger gauge symmetry (with or without SUSY) holds in the bulk. The symmetry is reduced in the 4 dimensional brane, where the physics that we observe is located, as an effect of symmetry breaking induced geometrically by suitable boundary conditions. There are models where SUSY, valid in $n>4$ dimensions is broken by boundary conditions \cite{ant}, in particular the model of ref.\cite{bar}, where the mass of the Higgs is computable and can be estimated with good accuracy. Then there are "Higgsless  models" where it is the SM electroweak gauge symmetry which is broken at the boundaries \cite{Hless}.  Or models where the Higgs is the 5th component of a gauge boson of an extended  symmetry valid in $n>4$ \cite{hoso}. In general all these alternative models for the Higgs mechanism face severe problems and constraints from electroweak precision tests \cite{BPR}. At the GUT scale, symmetry breaking by orbifolding can be applied to obtain a reformulation of SUSY GUT's where many problematic features of ordinary GUT's (e.g. a baroque Higgs sector, the doublet-triplet splitting problem, fast proton decay etc) are improved \cite{Kaw}, \cite{FeAa}.

\section{LITTLE HIGGS MODELS}

In "little Higgs" models the symmetry of the SM is extended to a suitable global group G that also contains some
gauge enlargement of $SU(2)\bigotimes U(1)$, for example $G\supset [SU(2)\bigotimes U(1)]^2\supset SU(2)\bigotimes
U(1)$. The Higgs particle is a pseudo-Goldstone boson of G that only takes mass at 2-loop level, because two distinct
symmetries must be simultaneously broken for it to take mass,  which requires the action of two different couplings in
the same diagram. Then in the relation
between
$\delta m_h^2$ and
$\Lambda^2$ there is an additional coupling and an additional loop factor that allow for a bigger separation between the Higgs
mass and the cut-off. Typically, in these models one has one or more Higgs doublets at $m_h\sim~0.2~TeV$, and a cut-off at
$\Lambda\sim~10~TeV$. The top loop quadratic cut-off dependence is partially canceled, in a natural way guaranteed by the
symmetries of the model, by a new coloured, charge-2/3, vectorial quark $\chi$ of mass around $1~TeV$ (a fermion not a scalar
like the s-top of SUSY models). Certainly these models involve a remarkable level of group theoretic virtuosity. However, in
the simplest versions one is faced with problems with precision tests of the SM \cite{prob}. Even with
vectorlike new fermions, large corrections to the epsilon parameters \cite{eps} arise from exchanges of the new gauge bosons
$W'$ and $Z'$ (due to lack of custodial $SU(2)$ symmetry). In order to comply with these constraints the cut-off must be
pushed towards large energy and the amount of fine tuning needed to keep the Higgs light is still quite large.
Probably these bad features can be fixed by some suitable complication of the model (see for example, \cite{Ch}). But, in my opinion, the real limit of
this approach is that it only offers a postponement of the main problem by a few TeV, paid by a complete loss of
predictivity at higher energies. In particular all connections to GUT's are lost. An interesting model that combines the idea of the Higgs as a Goldstone boson and warped extra dimensions was proposed and studied in refs.\cite{con}.

\section{DARK MATTER AND DARK ENERGY}

We know by now \cite{WMAP} that  the  Universe is flat and most of it is not made up of known forms of matter: $\Omega_{tot} \sim 1$, $\Omega_{baryonic} \sim 0.044$, $\Omega_{matter} \sim 0.3$, where $\Omega$ is the ratio of the density to the critical density. Most is Dark Matter (DM) and Dark Energy (DE) with $\Omega_{\Lambda} \sim 0.7$. We also know that most of DM must be cold (non relativistic at freeze-out) and that significant fractions of hot DM are excluded. Neutrinos are hot DM (because they are ultrarelativistic at freeze-out) and indeed are not much cosmo-relevant: $\Omega_{\nu} \lappeq 0.015$. Identification of DM is a task of enormous importance for both particle physics and cosmology. If really neutralinos are the main component of DM they will be discovered at the LHC and this will be a great service of particle physics to cosmology. More in general, the LHC is sensitive to a large variety of WIMP's (Weekly Interacting Massive Particles). WIMP's with masses in the 10 GeV-1TeV range with typical electroweak crosssections contribute to $\Omega$ terms of $o(1)$.
Also, these results
on cosmological parameters have shown that vacuum energy accounts
for about 2/3 of the critical density: $\Omega_{\Lambda}\sim 0.7$, Translated into familiar units this means for the energy
density $\rho_{\Lambda}\sim (2~10^{-3}~eV)^4$ or $(0.1~mm)^{-4}$. It is really interesting (and not at all understood)
that $\rho_{\Lambda}^{1/4}\sim \Lambda_{EW}^2/M_{Pl}$ (close to the range of neutrino masses). It is well known that in
field theory we expect $\rho_{\Lambda}\sim \Lambda_{cutoff}^4$. If the cut off is set at $M_{Pl}$ or even at $0(1~TeV)$
there would an enormous mismatch. In exact SUSY $\rho_{\Lambda}=0$, but SUSY is broken and in presence of breaking 
$\rho_{\Lambda}^{1/4}$ is in general not smaller than the typical SUSY multiplet splitting. Another closely related
problem is "why now?": the time evolution of the matter or radiation density is quite rapid, while the density for a
cosmological constant term would be flat. If so, then how comes that precisely now the two density sources are
comparable? This suggests that the vacuum energy is not a cosmological constant term, buth rather the vacuum expectation
value of some field (quintessence) and that the "why now?" problem is solved by some dynamical coupling of the quintessence field with gauge singlet fields (perhaps RH neutrinos) \cite{qui}. 

Clearly the cosmological constant problem poses a big question mark on the relevance of naturalness as a relevant criterion also for the hierarchy problem: how we can trust that we need new physics close to the weak scale out of naturalness if we have no idea on the solution of the cosmological constant huge naturalness problem? The common answer is that the hierarchy problem is formulated within a well defined field theory context while the cosmological constant problem makes only sense within a theory of quantum gravity, that there could be modification of gravity at the sub-eV scale, that the vacuum energy could flow in extra dimensions or in different Universes and so on. At the other extreme is the possibility that naturalness is misleading. Weinberg \cite{We} has pointed out that the observed order of magnitude of $\Lambda$ can be successfully reproduced as near the maximal value necessary to allow galaxy formation in the Universe. In a scenario where new Universes are continuously produced we might be living in a very special one (largely fine-tuned) but the only one to allow the development of an observer (anthropic principle). One might then argue that the same could in principle be true also for the Higgs sector. Recently it was suggested \cite{AHD} to abandon the no-fine-tuning assumption for the electro-weak theory, but require correct coupling unification, presence of dark matter with weak couplings and a single scale of evolution from the EW to the GUT scale. A "split SUSY" model arises as a solution with a fine-tuned light Higgs and all SUSY particles heavy except for gauginos, higgsinos and neutralinos, protected by chiral symmetry. But, then, we could also have a two-scale non-SUSY GUT with axions as dark matter. In conclusion, it is clear that naturalness can be a good heuristic principle but you cannot prove its necessity. The anthropic approach to the hierarchy problem is discussed in ref.s \cite{anto}.

\section{SUMMARY AND CONCLUSION}

Supersymmetry remains the standard way beyond the SM. What is unique to SUSY, beyond leading to a set of consistent and
completely formulated models, as, for example, the MSSM, is that this theory can potentially work up to the GUT energy scale.
In this respect it is the most ambitious model because it describes a computable framework that could be valid all the way
up to the vicinity of the Planck mass. The SUSY models are perfectly compatible with GUT's and are actually quantitatively
supported by coupling unification and also by what we have recently learned on neutrino masses. All other main ideas for going
beyond the SM do not share this synthesis with GUT's. The SUSY way is testable, for example at the LHC, and the issue
of its validity will be decided by experiment. It is true that we could have expected the first signals of SUSY already at
LEP, based on naturality arguments applied to the most minimal models (for example, those with gaugino universality at
asymptotic scales). The absence of signals has stimulated the development of new ideas like those of large extra dimensions
and "little Higgs" models. These ideas are very interesting and provide an important reference for the preparation of LHC
experiments. Models along these new ideas are not so completely formulated and studied as for SUSY and no well defined and
realistic baseline has sofar emerged. But it is well possible that they might represent at least a part of the truth and it
is very important to continue the exploration of new ways beyond the SM. New input from experiment is badly needed, so we all look forward to the start of the LHC.

\begin{acknowledgments}
I wishe to most warmly thank the Organizers of the SSI for their kind invitation and hospitality. I am also grateful to Dr. Paolo Gambino for providing me with an update of Figs. 3 and 4 and to Dr. Martin Grunewald for ref.\cite{gru}. 
\end{acknowledgments}


\begin{thebibliography}{99}   
\bibitem{ICHEP'06} D. Wood, Proceedings of ICHEP'06, Moscow, July 2006.
\bibitem{ewwg} The LEP Electroweak Working Group, http://lepewwg.web.cern.ch/LEPEWWG/
\bibitem{zzi} N. Cabibbo , L. Maiani, G. Parisi and R. Petronzio, {\it Nucl. Phys.} {\bf B158},295 (1979).
\bibitem{zzii} M. Sher, {\it Phys. Rep.} {\bf 179}, 273 (1989);  {\it Phys. Lett.}
{\bf B317}, 159 (1993).
\bibitem{aaiiii} G. Altarelli and G. Isidori, {\it Phys. Lett.} {\bf B337}, 141 (1994); J.A. Casas, J.R. Espinosa and M. Quir\'os,  {\it Phys. Lett.}
{\bf B342}, 171 (1995); J.A. Casas et al., {\it Nucl. Phys.} {\bf B436}, 3 (1995);  E{\bf B439}, 466 (1995); M. Carena and C.E.M. Wagner, {\it Nucl. Phys.} {\bf B452}, 45 (1995).
\bibitem{isi} G. Isidori, G. Ridolfi, A. Strumia;  {\it Nucl. Phys.} {\bf B609},387 (2001). 
\bibitem{eeiiii} See, for example, M. Lindner,  {\it Z. Phys.} {\bf 31}, 295 (1986);T. Hambye and K. Riesselmann, {\it Phys. Rev.} {\bf D55}, 7255 (1997).
\bibitem{AG} G.Altarelli and M. Grunewald, {\it Phys. Rep.} {\bf 403}, 189 (2004);  [arXiv:hep-ph0404165].
\bibitem{AC} T. Appelquist and J. Carazzone,  {\it Phys. Rev.}{\bf D11}, 2856 (1975).
\bibitem{P-Gambino} P.~Gambino, Int.\ J.\ Mod.\ Phys.\ A {\bf 19} (2004) 808,
 [arXiv:hep-ph 0311257].
 \bibitem{CTW} D. Choudhury, T.M.P. Tait and C.E.M. Wagner, Phys.Rev.D65,053002 (2002),
[ArXiv:hep-ph 0109097]; see also the recent discussion in K. Agashe, R. Contino, L. Da Rold and A. Pomarol, {\it Phys. Lett.}{\bf B641},62 (2006);
[ArXiv:hep-ph 0605341].
\bibitem{gru} M. Grunewald, private communication.
\bibitem{cha}
M.~S.~Chanowitz,  Phys. Rev. D 66 (2002) 073002, [ArXiv:hep-ph0207123];
\bibitem{acggr} G.~Altarelli, F.~Caravaglios, G.F.~Giudice, P.~Gambino and G.~Ridolfi, JHEP 0106:018, 2001,
[ArXiv:hep-ph 0106029]. 
\bibitem{eps} G.~Altarelli, R.~Barbieri and F.~Caravaglios,Int.\ J.\ Mod.\ Phys.\ A {\bf 13} {\bf A13}, 1031 (1998),  [ArXiv:hep-ph 9712368]. 
\bibitem{bar06} R. Barbieri, L. J. Hall and V. S. Rychkov
{\it Phys. Rev.} D74, 015007  (2006), [ArXiv:hep-ph 0603188].
\bibitem{CKMfitter} CKM Fitter group, http://ckmfitter.in2p3.fr/ and UT Fit group, http://utfit.roma1.infn.it/
 \bibitem{fle} For a recent review, see, for example, P. Ball, R. Fleischer, hep-ph 0604249.
\bibitem{alfe} For a review see, for example, G. Altarelli and F. Feruglio, {\it New J.Phys.}\textbf{6}106 (2004), [ArXiv:hep-ph 0405048].
\bibitem{buch} For a recent review, see, for example, W. Buchmuller, R.D. Peccei and T. Yanagida,{\it  Ann.Rev.Nucl.Part.Sci}.\textbf{55},311(2005),hep-ph 0502169.
\bibitem{BNL} G.W. Bennett et al., {\it Phys. Rev.}{\bf D73}, 072003 (2006);
[ArXiv:hep-ex 0602035] and refs. therein.
\bibitem{eid} S. Eidelman, Proceedings of ICHEP'06, Moscow, July 2006.
\bibitem{pas} For a review of the theory see, for example, M. Passera, {\it J.Phys.}G {\bf 31},R75 (2005); [ArXiv:hep-ph 0411168]
\bibitem{BS} R. Barbieri and A. Strumia, hep-ph 0007265.
\bibitem{L-C} K. Lane, hep-ph 0202255; R.S. Chivukula, hep-ph 0011264.
\bibitem{Martin} For a recent introduction see, for example, S. P. Martin, hep-ph 9709356. 
\bibitem{nmssm}H.~P.~Nilles, M.~Srednicki and D.~Wyler,
Phys.\ Lett.\ B \textbf{120}, 346(1983);
J.~P.~Derendinger and C.~A.~Savoy,
Nucl.\ Phys.\ B \textbf{237},  307(1984);
M.~Drees,
Int.\ J.\ Mod.\ Phys.\ A\textbf{4}, 3635 (1989);
J.~R.~Ellis, J.~F.~Gunion, H.~E.~Haber, L.~Roszkowski and F.~Zwirner,
Phys.\ Rev.\ D \textbf{39},  844 (1989);
T.~Elliott, S.~F.~King and P.~L.~White,
Phys.\ Lett.\ B \textbf{314}, 56 (1993)
hep-ph 9305282;
Phys.\ Rev.\ D \textbf{49}, 2435 (1994)
hep-ph 9308309;
U.~Ellwanger, M.~Rausch de Traubenberg and C.~A.~Savoy,
Phys.\ Lett.\ B \textbf{315}, 331 (1993)
hep-ph 9307322;
B.~R.~Kim, A.~Stephan and S.~K.~Oh,
Phys.\ Lett.\ B \textbf{336}  200 (1994).
\bibitem{barbie} ,  R. Barbieri, L. J. Hall, J. Nomura and V. S. Rychkov, hep-ph0607332.
\bibitem{gau} M. Dine and A. E. Nelson,{\it Phys. Rev.} \textbf{D48},  1277(1993);
M. Dine, A. E. Nelson and Y. Shirman,{\it Phys. Rev.}\textbf{D51}, 1362(1995);
G.F. Giudice and R. Rattazzi, {\it Phys. Rept.}\textbf{322}, 419 (1999).
\bibitem{ano} L. Randall and R. Sundrum, {\it Nucl. Phys.}\textbf{B557}, 79 (1999);
G.F. Giudice et al, JHEP\textbf{9812}, 027 (1998).
\bibitem{LP} P. Langacker and N. Polonsky, {\it Phys. Rev.}\textbf{D52}, 3081 (1995).
\bibitem{AFM} G. Altarelli, F. Feruglio and I. Masina, JHEP \textbf{0011},040 (2000).
\bibitem{kane} G. Kane et al, {\it Phys. Lett.}\textbf{ B551},146 (2003).
\bibitem{Jo} For a review and a list of refs., see, for example, J. Hewett and M. Spiropulu, hep-ph 0205196.
\bibitem{schm} For a review and a list of refs., see, for example, M. Schmaltz, hep-ph 0210415. 
\bibitem{RS} L. Randall and R. Sundrum, {\it Phys. Rev. Lett.}\textbf{ 83}, 3370 (1999), \textbf{83}, 4690 (1999).
\bibitem{FeAa} For a recent review, see for example, R. Rattazzi, hep-ph 0607055.
\bibitem{GW} W.D. Goldberger and M. B. Wise, Phys. Rev. Letters \textbf{83}, 4922 (1999), hep-ph 9907447.
\bibitem{ant} I. Antoniadis, C. Munoz and M. Quiros, {\it Nucl.Phys.}\textbf{B397},  515 (1993);
A. Pomarol and M. Quiros, Phys. Lett.\textbf{B438},  255 (1998).
\bibitem{bar} R. Barbieri, L. Hall and Y. Nomura, {\it Nucl.Phys.}\textbf{B624}, 63 (2002); R.Barbieri, G. Marandella and M. Papucci, hep-ph 0205280, hep-ph 0305044,  and refs therein.
\bibitem{Hless} see for example, C. Csaki et al, hep-ph 0305237 , hep-ph 0308038, hep-ph 0310355; S. Gabriel, S. Nandi and G. Seidl, hep-ph 0406020 and refs therein; R. Chivukula et al, hep-ph 0607124.
\bibitem{hoso}  see for example,  C.A. Scrucca, M. Serone and L. Silvestrini, hep-ph 0304220 and refs therein.
\bibitem{BPR} R. Barbieri, A. Pomarol and R. Rattazzi,  hep-ph 0310285.
\bibitem{Kaw} Y. Kawamura, {\it Progr. Theor. Phys.} \textbf{105}, 999 (2001).
\bibitem{prob} J. L. Hewett, F. J. Petriello, T. G. Rizzo, hep-ph 0211218;
C. Csaki et al, hep-ph 0211124, hep-ph 0303236.
\bibitem{Ch} H-C. Cheng and I. Low, hep-ph 0405243; J. Hubisz et al, hep-ph 0506042.
\bibitem{con} K Agashe, R. Contino, A. Pomarol, {\it Nucl.Phys.} \textbf{B719},165 (2005), hep-ph 0412089]; K. Agashe, R. Contino, {\it Nucl.Phys.}\textbf{B742},59 (2006), hep-ph 0510164]; K. Agashe, R. Contino, L. Da Rold, A. Pomarol, hep-ph 0605341].
\bibitem{tu} For orientation, see, for example, M. Turner, astro-ph 0207297.
\bibitem{WMAP} The WMAP Collaboration, D. N. Spergel et al, astro-ph 0302209.
\bibitem{qui} See, for example, R. Fardon, A.E. Nelson and N. Weiner, JCAP 0410 005 (2004), astro-ph 0309800; R. Barbieri, L. J. Hall, S. J. Oliver and A. Strumia, hep-ph 0505124.
\bibitem{We} S. Weinberg, {\it Phys. Rev. Lett.}\textbf{59},  2607 (1987). 
\bibitem{AHD} N.Arkani-Hamed and S. Dimopoulos, hep-th 0405159; G. Giudice and A. Romanino, hep-ph 0406088.
\bibitem{anto} N. Arkani-Hamed, S. Dimopoulos, S. Kachru, hep-ph 0501082;
G. Giudice, R. Rattazzi, hep-ph 0606105
\end{thebibliography}
\end{document}